\documentstyle[epsf,aps]{revtex}

\input epsf
\begin{document}


%
\vfill
\title{\Large\bf
The weak-field expansion for processes in a homogeneous background
magnetic field}

\vfill 
\author{Tzuu-Kang Chyi$^a$, Chien-Wen Hwang$^a$, W. F.  
Kao$^a$\thanks{email:wfgore@cc.nctu.edu.tw},
Guey-Lin Lin$^a$\thanks{email:glin@cc.nctu.edu.tw}, Kin-Wang Ng$^b$, Jie-Jun Tseng$^a$} %
\address{$^a$\rm Institute of Physics, National Chiao-Tung University,
Hsinchu, Taiwan \rm}
\address{$^b$\rm Institute of Physics, Academia Sinica, Taipei, Taiwan}
\date{\today}
\vfill
\maketitle


\begin{abstract}
The weak-field expansion of the charged fermion propagator
under a uniform magnetic field is studied.
Starting from Schwinger's proper-time representation, we
express the charged fermion propagator as an infinite series
corresponding to different Landau levels. This infinite series
is then reorganized according to the powers of the external field
strength $B$. For illustration, we apply this
expansion to  $\gamma\to \nu\bar{\nu}$ and $\nu\to \nu\gamma$ decays, which
involve
charged fermions in the internal loop. The leading and subleading magnetic-field
effects to the above processes are computed. 

\end{abstract}
%
%
%
%
\pagestyle{plain}


\section{Introduction}

Particle reactions taking place in the early universe or
astrophysical environments are often affected
by the background magnetic field
or the excitations in the medium \cite{REV}.
A typical example
is the modification of neutrino index of refraction in the early
universe
or the supernova \cite{refr}. There one needs to compute
the neutrino self-energy in the medium or
the background electromagnetic field or both. The neutrino index of refraction is
then extracted from the modified dispersion relation of the neutrino.  Another example is the
plasmon decay $\gamma^*\to \nu\bar{\nu}$\cite{REV} where the decaying photon
acquires an
effective mass through the effects of the medium or the background
magnetic field. With such an effective mass, the above decay is kinematically
permissible. Furthermore, the 
behaviors of electron propagators occurring in the internal loop of the
above decay are also affected by the medium or the magnetic-field. 
This also leads to a modification to the plasmon decay amplitude. 
Finally, a more recent example is the enhancement
of neutrino-photon scatterings due to the background magnetic field \cite{Sha,CTAS}.
At the lowest order in the weak interaction, it is
known that the amplitude for
$\gamma\gamma \to \nu\bar{\nu}$ is proportional to the neutrino mass \cite
{MGM}. Hence
the resulting scattering cross section is rather suppressed. On the other hand,
the presence of the background magnetic field alters the structures of
internal electron propagators, such that $ \gamma\gamma \to \nu\bar{\nu} $
is non-vanishing at $O(G_F)$ even in the massless limit of neutrinos.
Specifically,
the $\gamma\gamma \to \nu\bar{\nu}$
cross section is enhanced by a factor
$(m_W/m)^4(B/B_c)^2$ due to a background magnetic field $B$\cite{Sha,CTAS},
where $m_W$ and
$m$ are the masses of $W$ boson and electron respectively; $B_c\equiv
m^2/e$ is the critical magnetic field.

In the above processes, the relevant magnetic-field
strengths are often smaller than the critical value $B_c$.  Therefore
it is
appropriate to expand the decay width, cross section or other physical
quantities
in powers of $B/B_c$. In the literature, such an expansion is
usually performed after the relevant amplitude is obtained \cite{TSAI}.
For a more complicated process, it is not always
convenient to do so since the amplitude to be expanded may be
very cumbersome. In this article, we shall propose a more straightforward
weak-field expansion, which is performed directly on the charged fermion
propagator participating in the process. With the charged fermion
propagators expanded, the physical
amplitude can be easily expressed in powers of $B/B_c$.
To perform such an expansion on propagators, we shall begin with
Schwinger's proper-time representation for a charged fermion propagator
under a uniform background magnetic field\cite{Schw}. It is useful to
realize that Schwinger's representation can be recast into a series
expansion in terms of Landau levels \cite{chodos}. In the weak field limit $B\ll B_c$,
we shall demonstrate that one can reorganize the infinite series in powers of
the field strength $B$. This is the expansion we are after.

This article is organized as follows: In Sec. II, we will review Schwinger's
derivation of charged fermion propagator in a homogeneous background
magnetic field.
Since the convention used by Schwinger differs
from the currently popular convention, we shall repeat
some relevant details of the derivation for clarification.
We shall also illustrate how to rewrite Schwinger's result as an infinite
series where each term is associated with
specific Landau levels \cite{chodos}. In the weak-field limit,
we shall demonstrate how to rearrange the above series in powers of the
magnetic-field strength $B$. Finally, some technical issues relevant to
the phase factor in Schwinger's proper-time representation will be
discussed in this section. In Sec. III, 
we begin with a briefly discussion on our earlier work\cite{CTAS}, 
where
the weak-field expansion technique is applied to 
$\gamma\gamma\to \nu\bar{\nu}$ and its
crossed processes in a background magnetic field\cite{Sha,DR}.
To further illustrate the technique of weak-field expansion, we also 
calculate the decay rates of $\gamma\to
\nu\bar{\nu}$ and the neutrino Cherenkov process 
$\nu\to \nu\gamma$
in a background magnetic field. Our results will be compared to previous
calculations which are performed using exact 
charged-fermion propagators in the background
magnetic field\cite{GN,SK,DMH}.     
A few concluding remarks are presented in Sec. IV.


\section{Charged-fermion Propagator in a Homogeneous Background Magnetic
Field.}


\subsection{The Exact Propagator Solution}
The Green's function $G(x,x')$ of the Dirac field in the presence of a gauge
field
$A_{\mu}$ satisfies the following equation
\begin{equation}
\left(i\not{\!\partial}+e\not{\!\!A}-m\right)G(x,x')=\delta(x-x'),
\label{green} \end{equation}
where $\delta(x-x')$ is the Dirac's delta function and $m$ stands for the mass of the 
Dirac field. We will follow the technique
employed in Schwinger's paper \cite{Schw} which regards $G(x,x')$ as the matrix element
of an
operator $G$, namely $G(x,x')=\langle x'\vert G \vert x\rangle$. Therefore,
Eq. (\ref{green}) may be written as

\begin{equation}
\left(\not{\!\Pi} -m\right)G=1, \label{mgreen}
\end{equation}
with $\Pi_{\mu}=P_{\mu}+eA_{\mu}$ denoting the conjugated momentum, which
obeys the following commutation relations
\begin{eqnarray}
\left[\Pi_{\mu}, x_{\nu}\right]&=&i g_{\mu\nu}, \\
\left[\Pi_{\mu}, \Pi_{\nu}\right]&=& ieF_{\mu\nu},
\end{eqnarray}
with $F_{\mu\nu}\equiv \partial_{\mu}A_{\nu}-\partial_{\nu} A_{\mu}$ denoting
the field-strength tensors of the gauge field. Eq. (\ref{mgreen}) can be
formally  solved by writing
\begin{equation}
G={1\over \not{\!\Pi}-m }
= -i\int_0^{\infty}ds \left(\not{\!\Pi} +m\right) \exp[{-i(m^2-{\not{\!\Pi}}^2
)s}].
\end{equation}
This integral representation for $G$ implies that
\begin{equation}
G(x,x')=-i\int_0^{\infty}ds e^{-im^2s}\langle x'\vert (\not{\!\Pi} +m)U(s)
\vert x\rangle 
\end{equation}
where $U(s)=e^{-iHs}$ with $H\equiv -(\not{\!\!\Pi})^2=-\Pi^2 -{1\over
2}e\sigma_{\mu\nu}F^{\mu\nu}$. We observe that $U(s)$ can be viewed as the
unitary time-evolution
factor if one takes $H$ as the effective Hamiltonian that evolves the state
$\vert x \rangle$ according to
\begin{equation}
\vert x(s)\rangle=U(s)\vert x(0)\rangle ,
\end{equation}
where $s$ is the proper time variable. One can now rewrite $G(x,x')$
as
\begin{equation}
G(x,x')=-i\int_0^{\infty}ds e^{-im^2s}\left[ \gamma^{\mu}\langle x'(0)\vert
\Pi_{\mu}(s)\vert x(s)\rangle +m\langle x'(0)\vert x(s)\rangle
\right] ,
\label{Gss}
\end{equation}
where we have assumed $\Pi_{\mu}(s)$ operates on $\vert x(s)\rangle$ and
$\Pi_{\mu}(0)$ operates
on $\vert x(0)\rangle$. We note that
the operators $x_{\mu}$ and $\Pi_{\mu}$ satisfy
\begin{eqnarray}
{dx_{\mu}\over ds}&=&-i\left[x_{\mu}, H\right]=2\Pi_{\mu},\nonumber \\
{d\Pi_{\mu}\over ds}&=&-i\left[\Pi_{\mu}, H\right]=-2eF_{\mu\nu}\Pi^{\nu},
\label{OPEQ} \end{eqnarray}
for a constant field strength $F_{\mu\nu}$. In the matrix notation,
we may write
${dx/
ds}=2\Pi$, and ${d\Pi/ ds}=-2eF\Pi$. Furthermore the transformation
function $\langle x'(0)\vert x(s)\rangle$ can be characterized by the following
equations:
\begin{eqnarray}
i\partial_s\langle x'(0)\vert x(s)\rangle&=&\langle x'(0)\vert H\vert
x(s)\rangle,\nonumber \\
\left(i\partial_{\mu}+eA_{\mu}(x)\right)\langle x'(0)\vert x(s)\rangle
&=&\langle x'(0)\vert \Pi_{\mu}(s)\vert x(s)\rangle,\nonumber \\
\left(-i\partial^{\prime}_{\mu}+eA_{\mu}(x')\right) \langle x'(0)\vert
x(s)\rangle &=&\langle x'(0)\vert \Pi_{\mu}(0)\vert x(s)\rangle, \label{TREQ}
\end{eqnarray}
with the boundary condition: $\langle x'(0)\vert x(s)\rangle \to \delta^4(x-x')$
as $s\to 0$.
To evaluate Eq. (\ref{Gss}), we first solve
Eq. (\ref{OPEQ}) and obtain
\begin{eqnarray}
\Pi(s)&=&e^{-2eFs}\Pi(0),\nonumber \\
x(s)-x(0)&=& \left(1-e^{-2eFs}\right)\left(eF\right)^{-1} \Pi(0).
\label{pixs}
\end{eqnarray}
This solution implies
\begin{eqnarray}
\Pi^2&\equiv& -H-{1\over 2}e\sigma_{\mu\nu}F^{\mu\nu}\nonumber \\
& =&
(x(s)-x(0)) K (x(s)-x(0)), \nonumber \\
\left[ x_\mu(s), x_\nu(0) \right] & =&
i (1 -e^{-2eFs}) (eF)^{-1} ,
\end{eqnarray}
where $K \equiv {1 \over 4} (eF)^2 \sinh^{-2}eFs$.
Therefore, one has
\begin{equation}
\langle x'(0)\vert H\vert
x(s)\rangle =- {1 \over 2}e \sigma F - (x-x')K(x-x')- {i \over 2} {\rm tr} (eF
\coth eFs) .
\end{equation}
With this result, one can
solve the first equation in (\ref{TREQ}), which gives
\begin{eqnarray}
\langle x'(0)\vert x(s) \rangle &=& C(x,x')s^{-2} \exp\left[-{1\over 2}{\rm
tr}\ln  [(eFs)^{-1}\sinh(eFs)]\right]\nonumber \\
&\times&\exp\left[-{i\over 4}(x-x')eF\coth(eFs)(x-x')+{i\over
2}e\sigma_{\mu\nu} F^{\mu\nu}s\right].\label{212}
\end{eqnarray}
The factor $C(x,x')$ can be determined by substituting Eq. (\ref{212})
into the second and third equations in (\ref{TREQ}).  Since the r.h.s. of
these two equations are given by
\begin{eqnarray}
\langle x'(0)\vert \Pi(s)\vert x(s)\rangle &=& {1\over 2}
\left[eF\coth(eFs)-eF\right](x- x')\langle x'(0)\vert x(s)\rangle ,\nonumber \\
\langle x'(0)\vert \Pi(0)\vert x(s)\rangle &=& {1\over 2}
\left[eF\coth(eFs)+eF\right](x-x')\langle x'(0)\vert x(s)\rangle , \label{ccc}
\end{eqnarray}
one then arrives at
\begin{eqnarray}
\left[i\partial_{\mu}+eA_{\mu}(x)-{1\over 2}eF_{\mu\nu}(x'-x)^{\nu}
\right]C(x,x')&=&0 , \nonumber\\
\left[-i\partial^{\prime}_{\mu}+eA_{\mu}(x')+{1\over 2}
eF_{\mu\nu}(x'-x)^{\nu} \right]C(x,x')&=&0.
\end{eqnarray}
Therefore $C(x,x')$ is found to be
\begin{eqnarray}
C(x,x')&=&C'(x')\exp\left[ie\int_{x'}^x d\xi^{\mu}\left( A_{\mu}+{1\over
2}F_{\mu\nu}(\xi-x')^{\nu}\right)\right]\nonumber \\
&=&C(x)\exp\left[ie\int_{x'}^x d\xi^{\mu}\left( A_{\mu}+{1\over
2}F_{\mu\nu}(\xi-x)^{\nu}\right)\right]. 
\end{eqnarray}
Here $C'(x')$ and $C(x)$ denote integration constants in $x'$ and $x$
respectively.
Note that the integral $A_{\mu}+{1\over 2}F_{\mu\nu}(\xi-x')^{\nu}$ is a total
derivative in the presence of a homogeneous field if the first
homology group of the space-time $M$ is trivial, i.e. $H_1(M)=0$ \cite{topo}. Hence
the phase factor is independent of the integration
path connecting $x$ and $x'$. One can further show that $C(x')=C'(x)$.
Therefore
$C(x')$ or $C'(x)$ has to be a constant independent of $x$ and $x'$. This
constant can be
determined by applying the boundary condition $\langle x(s)\vert
x'(0)\rangle\to \delta^4(x-x')$ as $s\to 0$. One obtains
\begin{equation}
C=-i(4\pi)^{-2} \label{cpi}
\end{equation}
with the help of the identity
\begin{equation}
\int^{\infty}_{-\infty}e^{ia^2x^2}dx = \sqrt{i \pi \over a^2} .
\end{equation}
From Eqs. (\ref{Gss}), (\ref{212}), (\ref{ccc}) and (\ref{cpi}), one arrives at
\begin{eqnarray}
G(x,x') = \Phi (x,x') {\cal G}(x,x'),
\end{eqnarray} 
where
\begin{eqnarray}
{\cal G}(x,x')&\equiv&-(4\pi)^{-2}\int_0^{\infty}{ds\over s^2}
\left[m+{1\over 2}\gamma \cdot \left(eF\coth(eFs)-eF\right)(x-x') \right]
\exp (-im^2s + {i\over 2}e\sigma_{\mu\nu}F^{\mu\nu}s)
\nonumber \\
&\times& \exp\left[-{1\over 2}{\rm tr}\ln [(eFs)^{-1}\sinh(eFs)] -{i\over 4}(x-
x')\left(eF\coth(eFs)\,\,\right)\,\,(x-x')\right], \label{calGx}   \\
\Phi (x,x') &\equiv& \exp\left\{ie\int^x_{x'} d\xi^\mu \left[A_\mu +
{1\over{2}}F_{\mu\nu} (\xi-x')^{\nu}\right]\right\}. \label{phase}
\end{eqnarray} 
Note that the translation invariance is broken
by the phase factor $\Phi(x,x')$.
Note also that the phase factor $\Phi(x,x')$ vanishes if the path 
connecting $x$ and $x'$ is chosen to be
a straight-line. In addition, if
the background gauge field is a homogeneous magnetic
field such that $F_{12}=-
F_{21}=B$, one can show that
\begin{eqnarray}
\sigma_{\mu\nu}F^{\mu\nu} &=&  2F_{12} \sigma_3\equiv
2F_{12}\left(
\begin{array}{cc}
{\bf \sigma_3} & 0 \\
0 & {\bf \sigma_3}
\end{array}
\right), \nonumber \\
\exp[-{1\over 2}{\rm tr}\ln (\,F^{-1}\sinh \,F\,) \,\,]&=&
{B\over \sin B }, \nonumber \\
\gamma\left(F\coth F \,-F\right)x &=&
(\gamma\cdot x)_{\parallel}- {B\over \sin B}
(\gamma\cdot x)_{\bot} e^{iF_{12}\sigma_3}, \nonumber \\
x\left(F\coth F \right)x &=&
x_{\parallel}^2-B\cot B \,\, x_{\bot}^2,\label{rel}
\end{eqnarray}
with
$(a\cdot b)_{\parallel}=a^0b^0-a^3b^3$ and $(a\cdot
b)_{\bot}=a^1b^1+a^2b^2$
for arbitrary 4-vectors $a^{\mu}$ and $b^{\mu}$.
Hence $a^2_{\parallel}=a^0a^0-a^3a^3$, and $a^2_{\bot}=a^1a^1+a^2a^2$.
To simplify the notations, we shall denote $(\gamma\cdot p)_{\parallel (\bot)}$
as $\gamma\cdot p_{\parallel (\bot)}$.
From the relations in (\ref{rel}),
the propagator function ${\cal G}(x,x')$, which respects the
translation invariance, becomes
\begin{eqnarray}
{\cal G}(x)&=&-(4\pi)^{-2}\int_0^{\infty}{ds\over s^2}{eBs\over \sin(eBs)}
\exp(- im^2s +ieBs\sigma_3)\nonumber \\
&\times& \exp\left[-{i\over 4s}(x_{\parallel}^2-eBs\cot(eBs) \,\, x_{\bot}^2)\right]
\nonumber \\
&\times& \left[m+{1\over 2s}\left(\gamma\cdot x_{\parallel}- {eBs\over
\sin(eBs)}
\exp({-ieBs\sigma_3}) \,\, \gamma\cdot x_{\bot}\right)\right] .\label{G18}
\end{eqnarray}


\subsection{Weak Field Limit}
We find it is more convenient to cast (\ref{G18}) in the form \cite{TSAI}
\begin{eqnarray}
{\cal G}(x,x')=\int {d^4p\over {(2 \pi)^4}} e^{-ip(x-x')} {\cal G}(p)
\end{eqnarray} 
with
\begin{eqnarray}
{\cal G}(p)&=&\int d^4x  e^{ipx} {\cal G}(x)\nonumber \\
&=&-i\int^\infty_0 {ds\over {\cos(eBs)}}
\exp\left[-is\left(m^2-p^2_{\parallel}+
{\tan(eBs)\over{eBs}} p^2_{\bot}\right)\right]\nonumber \\
&&~~~~~~~\times \left[\exp({ieBs\sigma_3})(m+ \gamma\cdot p_{\parallel}\,)-
{ \gamma
\cdot p_{\bot} \over {\cos(eBs)}}\right] . \label{Gp}
\end{eqnarray} 
One can further show that
\begin{eqnarray}
{\cal G}(p)&=&-i \int^\infty_0 {ds\over{\cos(eBs)}}\exp\left[-
is\left(m^2-p^2_{\parallel}+ {\tan(eBs)\over{eBs}} p^2_{\bot}\right)\right]
\nonumber \\
&\times& \left[ \left[ \cos(eBs)+\gamma_1\gamma_2\sin(eBs) \right]
(m+\gamma\cdot
p_{\parallel})- {\gamma \cdot p_{\bot}\over {\cos(eBs)}}\right] \label{Gp2}
\end{eqnarray} 
when the following identities are applied:
\begin{eqnarray}
\exp({iz\sigma_3}) &=& \cos z{\bf I} +i \sin z \sigma_3, \\
\sigma_3 &\equiv& \left(
\begin{array}{cc}
{\bf \sigma_3} & 0\\
0 & {\bf \sigma_3}
\end{array}
\right)
= i\gamma_1\gamma_2.
\end{eqnarray} 
If we
define a new variable $v\equiv eBs$, then Eq. (\ref{Gp2}) can be rewritten as \cite{chodos}
\begin{equation}
{\cal G}(p)
\equiv -i \int_0^{\infty} dv \exp({-iv\rho}){1\over{eB}}\left[(m+\gamma \cdot
p_{\parallel}) I_1 +\gamma_1 \gamma_2  (m+\gamma\cdot p_{\parallel})
I_2-(\gamma \cdot p_{\bot}) I_3\right],\label{III}
\end{equation}
where
\begin{eqnarray}
I_1 &=&  \exp({-i\alpha \tan v}), \nonumber \\
I_2 &=&  \exp({-i\alpha \tan v}) \tan v, \nonumber \\
I_3 &=&  \exp({-i\alpha \tan v}) {1\over{\cos^2 v}},
\end{eqnarray}
with
$\rho \equiv (m^2-p^2_{\parallel})/eB$ and $\alpha \equiv  p^2_{\bot}/eB$.
Because $I_j (v) =I_j (v+n \pi)$ for $j=1,2,3$, we get
\begin{eqnarray}
\int_0^{\infty} dv \exp({-iv\rho}) I_j &=& \sum^{\infty}_{n=0}
\exp({-i\rho n \pi})
\int^\pi_0 dv \exp({-i\rho v}) I_j(v) \nonumber \\
&=& {1\over{1-e^{-i\rho \pi}}}\int^\pi_0 dv \exp({-i\rho v}) I_j \nonumber \\
&\equiv& {1\over{1-e^{-i\rho \pi}}} A_j. \label{IA}
\end{eqnarray}
It is sufficient to evaluate $A_1$ since the other
integrals are obtained using
\begin{eqnarray}
A_2&=&i{\partial \over{\partial \alpha}}A_1, \nonumber \\
A_3&=&{-i\over{\alpha}}(1-e^{-i\rho \pi})-{\rho\over{\alpha}} A_1. \label{AAA}
\end{eqnarray}
To evaluate $A_1\equiv\int^{\pi}_0 dv
\exp[{-i\alpha \tan v \,}] \exp({-i\rho v})$,
we rewrite
\begin{eqnarray}
\exp[{-i\alpha \tan v \,}] =
\exp\left[\alpha {-e^{-2iv}+1\over{-e^{-2iv}-1}}\right]
\label{ee}.
\end{eqnarray} 
The r.h.s. of this equation can be expanded using the
Laguerre polynomials. Specifically, the Laguerre polynomials
$L_n(x)$ are generated by the following generating function
\begin{eqnarray}
{\exp[{-xZ/(1-Z)}]\over{1-Z}}=\sum^\infty_{n=0} L_n(x) Z^n \label{Lag}
\end{eqnarray} 
for $|Z|\leq 1$.
Upon multiplying $Z$ on both sides of (\ref{Lag}) and subtracting
(\ref{Lag}), one arrives at
\begin{eqnarray}
\exp\left[{-xZ\over{1-Z}}\right]=\sum^\infty_{n=0} (L_n(x)-L_{n-1}(x)) Z^n
\label{LLag},
\end{eqnarray} 
where one sets $L_{-1}(x)=0$.
Using the identity
\begin{eqnarray}
\exp\left({x\over{2}} {Z+1\over{Z-1}}\right)=
\exp\left[{-{xZ\over {1-Z}}}\right]\cdot \exp\left(-{x\over2}\right)
\label{zzz}
\end{eqnarray} 
with the identifications $Z\equiv -e^{-2iv}$, $x\equiv 2\alpha$, and
combining Eqs. (\ref{IA}), (\ref{ee}), and
(\ref{LLag}),
one obtains
\begin{eqnarray}
A_1&=&\int^\pi_0 dv e^{-\alpha} \sum^\infty_{n=0}(L_n(2 \alpha)-L_{n-1} (2
\alpha)) \exp({-2inv}) (-1)^n \exp({-i\rho v})\nonumber \\
&=& e^{-\alpha} \sum^\infty_{n=0} C_n(2 \alpha) (-1)^n \int^{\pi}_0 dv
\exp[{-i(\rho+2n) v}] \nonumber \\
&=& -i e^{\alpha}(1-e^{-i\rho\pi}) \sum^\infty_{n=0} {(-1)^n C_n(2
\alpha)\over{\rho+2n}}. \label{AC}
\end{eqnarray} 
Using Eqs. (\ref{III}),
(\ref{AAA}) and (\ref{AC}),
one rewrites the
propagator function ${\cal G}(p)$ into a simple form \cite{chodos}
\begin{eqnarray}
i{\cal G}(p)=
\sum^\infty_{n=0} {-id_n(\alpha) D+d'_n(\alpha)
\bar D \over{p^2_L+2 neB}} +i{\gamma
\cdot p_{\bot}\over{p^2_\bot}}, \label{DDb} 
\end{eqnarray} 
where
$d_n(\alpha)\equiv (-1)^n e^{-\alpha}C_n(2\alpha)$,
$d'_n=\partial d_n/\partial \alpha, p^2_L=m^2-p^2_\parallel$, and
\begin{eqnarray}
D &=& (m+\gamma\cdot p_{\parallel})+\gamma\cdot p_{\bot} {m^2-p^2_{\parallel}
\over{p^2_{\bot}}},\nonumber \\
\bar D &=& \gamma_1 \gamma_2 (m+\gamma \cdot
p_{\parallel} ) . \label{DDe} \end{eqnarray} 

We note that,
in the limit of extreme field strength, i.e. $B\gg B_c$ or $B\ll B_c$,
only part of the terms in Eq. (\ref{DDb}) are relevant. In the strong
field limit $B\gg B_c$, only contributions from the lowest
Landau level $n=0$ need to be kept. For the weak field limit
$B\ll B_c$, we shall demonstrate that
the infinite series in Eq. (\ref{DDb}) may be
reorganized in powers of the magnetic field $B$. Therefore those terms with
lower powers of $B$ are more important in this limit.
To reorganize the series,
we first observe that
\begin{eqnarray}
\sum^\infty_{n=0} {-id_n D+d'_n \bar D \over{p^2_L+2 ne
B}}&=&{1\over{p^2_L}}\sum^\infty_{n=0} {-id_n D+d'_n \bar D
\over{1+{2neB\over{p^2_L}}}}\nonumber \\
&=&{1\over{p^2_L}}\sum^\infty_{n=0}\sum^\infty_{k=0}(-id_n D+d'_n \bar D)
\left(-2neB\over{p^2_L}\right)^k \nonumber \\
&=& \sum^\infty_{k=0} {1\over{p^2_L}} \left(-2eB\over{p^2_L}\right)^k \left(-
iD\sum^\infty_{n=0} n^k d_n(\alpha)+\bar D \sum^\infty_{n=0} n^k d'_n(
\alpha)\right). \label{1term}
\end{eqnarray} 
The infinite series $\sum^\infty_{n=0} n^k d_n(\alpha)$ and
$\sum^\infty_{n=0} n^k d'_n(\alpha)$ can be evaluated with the
the identity
\begin{eqnarray}
\sum^\infty_{n=0} d_n(\alpha) \exp({-2inv})=
\exp[{-i\alpha \tan v }],\label{dae}
\end{eqnarray} 
which follows from Eqs. (\ref{ee}), (\ref{LLag}) and (\ref{zzz}).
Let us proceed by taking a derivative $\partial /\partial v$
on both sides of (\ref{dae}).
This gives
\begin{eqnarray}
(-2i)^1 \sum^\infty_{n=0} n^1 d_n(\alpha)
\exp({-2inv})= {-i\alpha\over{\cos^2 v }}
\exp[{-i\alpha \tan v }].\nonumber
\end{eqnarray} 
Taking this derivative $k$ times, we find that
\begin{equation}
(-2i)^k \sum^\infty_{n=0} n^k d_n(\alpha) \exp({-2inv}) = \Bigg\{\Bigg({-
i\alpha\over{\cos^2 v \,}}\Bigg)^k+
O(\alpha^{k-1})
\Bigg\}\exp[{-i\alpha \tan v\,}]. \label{sum}
\end{equation}
To be more specific,
one can define $U(v)\equiv \exp[{-i\alpha \tan v }]$ following Eq. (\ref{dae}).
It can be shown that
$\partial_v U=FU$ with $F \equiv -i \alpha/\cos^2 v$. Hence one can show that
\begin{eqnarray}
\partial_v^k U &=& \sum_{l=0}^{k-1} C^{k-1}_l 
\partial_v^{k-l-1} F \partial_v^l U  \nonumber \\ 
&=& \left[  F^k +C^k_2 F^{k-2} 
\partial_vF + C^k_3 F^{k-3}
\partial_v^2F+ C^3_2 C^k_4 
F^{k-4} 
(\partial_v F)^2 \right] 
+ \kappa_3 (\alpha) + \kappa_4 (\alpha) +O(\alpha^{k-5}) \label{dkU} .
\end{eqnarray} 
Here $C^a_b \equiv a!/[b!(a-b)!]$ 
denotes the number of combinations of size $b$ from a collection of size $a$. 
In addition, $\kappa_3$ and $\kappa_4$ denote the third and fourth derivative
terms. They can be shown to be
\begin{eqnarray}
\kappa_3( \alpha) &=&
C^k_4 F^{k-4} \partial^3_vF +  C^k_5 C^5_2 F^{k-5} \partial_vF \partial_v^2F + C^6_2 C^k_6 
F^{k-6} 
(\partial_v F)^3  \nonumber \\ 
\kappa_4( \alpha) &=&
C^k_5 F^{k-5} \partial^4_vF +  C^k_6 C^6_4 F^{k-6} \partial_vF \partial_v^3F + C^5_2 C^k_6 
F^{k-6} (\partial^2_v F)^2  \nonumber \\ &&
+ C^k_7 C^7_4 C^3_2 F^{k-7} \partial^2_vF (\partial_vF)^2
+ C^k_8 C^7_3 C^3_2  F^{k-8} (\partial_v F)^4 
\end{eqnarray} 
Note that above formula for the expansion of $\partial^k_v U$ can either be proved
by method of induction or can be read off directly from the combinatorial factor in the 
the expansion of $(\partial_v + F)^k \cdot 1$ \cite{combine}.
It is worthy pointing out that $\partial_v^k F(v=0)=0$ for all 
odd number $k$ and the value of $\partial_v^k F(v=0)$ when $k$ is even can be computed directly.
For example one can show that 
$\partial_v^2F(v=0)=2F(v=0)$, and $\partial_v^4F(v=0)=16F(v=0)$. 
Hence the order of $\alpha^{k-2}$ and the order of $\alpha^{k-4}$ terms read
$\partial_v^k U (v=0) =2C^k_3 (-i \alpha)^{k-2} + [16 C^k_5 +40C^k_6 ] (-i \alpha)^{k-4}$. 
Similarly, one can also show that the order of $\alpha^{k-n}$ term for the
$\bar{D}$ term vanishes when $n$ is an even integer while $D$ term vanishes for all odd integer $n$.
Hence, by setting $v=0$ on both sides of Eq. (\ref{sum}), we obtain
\begin{eqnarray}
\sum^\infty_{n=0} n^k d_n(\alpha) &=& \left({\alpha\over{2}}\right)^k 
- { 1\over 2}C^k_3 \left( {\alpha \over 2} \right)^{k-2} 
+\left [ C^k_5 + { 5 \over 2} C^k_6 \right] \left( {\alpha \over 2} \right)^{k-4}
,\nonumber \\
\sum^\infty_{n=0} n^k d'_n(\alpha) &=&
{k\over{2}}\left({\alpha\over{2}}\right)^{k-1}
- { k-2 \over 4}C^k_3 \left( {\alpha \over 2} \right)^{k-3} 
+ {k-4 \over 2}
\left [ C^k_5 + { 5 \over 2} C^k_6 \right] \left( {\alpha \over 2} \right)^{k-5}
+O(\alpha^{k-6}). \label{2app}
\end{eqnarray}
Here we only keep terms to the order of $O(\alpha^{k-5})$. 
Since $\alpha=p_{\bot}^2/eB$, the leading terms
on the r.h.s. of the above equation give up to order of $O(e^3B^3)$
contributions to ${\cal G}(p)$, as can be seen from Eq. (\ref{1term}).
Precisely we have
\begin{eqnarray}
\sum^\infty_{n=0} {-id_n D+d'_n \bar D \over{p^2_L+2 ne B}} &=&
\sum^\infty_{k=0} {1\over{p^2_L}} \left(-2eB\over{p^2_L}\right)^k \Bigg\{-iD
\left[ \left({\alpha\over{2} }\right)^k 
- { 1\over 2}C^k_3 \left( {\alpha \over 2} \right)^{k-2} \right]
\nonumber  \\ 
&& +\bar D \left[ {k\over 2} 
\left( {\alpha\over 2} \right)^{k-1} - { k-2 \over 4}C^k_3 \left( {\alpha \over 2} 
\right)^{k-3} \right]
     \Bigg\} +  i{\cal G}_4 (p) \nonumber \\
&=& \sum^\infty_{k=0} {1\over{p^2_L}}\left[-iD\left({-
p^2_\bot\over{p^2_L}}\right)^k+\bar D
\left({-p^2_\bot\over{p^2_L}}\right)^{k-1}
\left(-k\over{p^2_L}\right) eB\right] \nonumber \\
&+& \sum^\infty_{k=0} {1\over{p^2_L}}
\left[ i2C^k_3 D  \left( {eB \over p_L^2} \right)^2 \left( 
{p^2_\bot\over{p^2_L}}\right)^{k-2}+2(k-2)C^k_3 \bar D \left( {eB\over{p^2_L}}\right)^3
\left({-p^2_\bot\over{p^2_L}}\right)^{k-3}
\right] + i{\cal G}_4 (p) \nonumber \\
&=& {-iD\over{p^2_L}} {1\over{1+{p^2_\bot \over{p^2_L}}}}-{\bar D
\over{(p^2_L)^2}} {1\over { \left( 1+{p^2_\bot \over{p^2_L}} \right)^2}}eB +i{\cal G}_2 (p) 
+i{\cal G}_4 (p) \nonumber \\
&=& {iD\over{p^2-m^2}}-{\bar D\over{(p^2 -m^2)^2}}eB +i{\cal G}_2(p) +i{\cal G}_4 (p). 
\label{1termf}
\end{eqnarray} 
Where $i{\cal G}_2(p)$ denotes terms of order $e^2B^2$ and $e^3B^3$
and $i{\cal G}_4 (p)$ denotes terms of order $e^4B^4$ and $e^5B^5$.
Therefore, by Eqs. (\ref{DDe}) and ({\ref{DDb}), we arrive at
\begin{eqnarray}
i {\cal G}(p) &=& {iD\over{p^2-m^2}}-{\bar D\over{(p^2 -m^2)^2}}eB+i{\gamma
\cdot p_\bot \over {p^2_\bot}} +i{\cal G}_2(p) +i{\cal G}_4 (p) \nonumber \\
&=& {i(\not{\! p}+m)\over{p^2-m^2}}-{\gamma_1\gamma_2(\gamma
\cdot p_\parallel+m)\over{(p^2-m^2)^2}} eB +i{\cal G}_2(p) +i{\cal G}_4 (p). 
\label{weakB}
\end{eqnarray} 
The first term of Eq. (\ref{weakB}) is just the electron propagator in
the vacuum, while the second term is its correction to $O(eB)$.
The corrections with higher powers in $eB$ can be calculated in a similar way.
For example, to evaluate the term ${\cal G}_2(p)$ and ${\cal G}_4(p)$,
we need to compute a few identities.
Note that $C^k_3=k(k-1)(k-2)/6$ hence one can show that
$\sum_{k=0}^\infty C^k_3 (-x)^{k-3}/6 =1/(1+x)^4$ for all $|x| <1$ from consecutive 
differentiating the identity 
$\sum_{k=0}^\infty (-x)^k =1/(1+x)$ which is valid for all $|x| <1$.
Similarly, one can also show that
$\sum_{k=0}^\infty (k-2)C^k_3(-x)^{k-3} =1/(1+x)^4-4x / (1+x)^5$ for all $|x| <1$.
Therefore, one can extract the $O(e^2B^2)$ and $O(e^3B^3)$ terms from the series expansion in Eq. 
(\ref{1term}). 
The result leads to the following contribution:
\begin{equation}
i{\cal G}_2(p)=-{2ie^2B^2p_{\bot}^2\over (p^2-m^2)^4}D
+2 e^3B^3 \left[ {1 \over (p^2-m^2)^4} + {4 p_{\bot}^2 \over (p^2-m^2)^5} \right] \bar D ,
\label{B23}
\end{equation}
where $D$ and $\bar D$ have been defined in Eq. (\ref{DDe}).
Similarly, one can show that
$\sum_{k=0}^\infty C^k_5 (-x)^{k-5} =1/(1+x)^6$, 
$\sum_{k=0}^\infty C^k_6 (-x)^{k-6} =1/(1+x)^7$, 
$\sum_{k=0}^\infty (k-4) C^k_5 (-x)^{k-5} =1/(1+x)^6 - 6x /(1+x)^7$ and 
$\sum_{k=0}^\infty (k-4) C^k_6 (-x)^{k-6} =2/(1+x)^7 - 7x /(1+x)^8$ 
for all $|x| <1$.
Hence the fourth and fifth order propagator $i{\cal G}_4 (p)$ can be shown to be
\begin{equation}
i{\cal G}_4(p)=-[8ie^4B^4] \left [ { 2 p_{\bot}^2 p_L^2 -3 (p_{\bot}^2)^2 \over (p^2-m^2)^7}
\right] D 
-8 e^5B^5 \left[ {15 ( p_{\bot}^2 )^2 -16 p_\bot^2 p_L^2 +2 p_L^2  
\over (p^2-m^2)^8} \right] \bar D.
\label{B45}
\end{equation}


\subsection{Phase Factor}
In this subsection, we discuss how to treat the
phase factor $\Phi(x,x')$ as defined in Eq. (\ref{phase}).
First, we
note that $\Phi(x,x')$ is reduced to
\begin{eqnarray}
\Phi(x,x')=\exp\left\{ie\int^x_{x'}d\xi^\mu A_\mu(\xi)\right\},
\end{eqnarray} 
if the integration path connecting $x$ and $x'$ is a
straight line. This choice of integration path is only
for convenience since the
integration in Eq. (\ref{phase}) is
path independent provided that the vector potential
$A_{\mu}(\xi)$ is non-singular.
Second, for a particular type of Coulomb gauge:
\begin{eqnarray}
A_0 (\xi) &=& 0, \nonumber \\
{\bf A} (\xi)&=& {B\over{2}}(x'_{2}-\xi_2,\xi_1-x'_{1},0),\nonumber
\end{eqnarray} 
the exponent $\int^x_{x'}d\xi^\mu A_\mu(\xi)$ vanishes,
hence $\Phi(x,\, x')=1$.
Therefore, by choosing the above Coulomb gauge, the phase factor $\Phi(x,x')$
in the electron Green's function can be disregarded.
Such a simplification is, however, no longer valid
for more complicated processes where
more than one electron propagators are involved in the process.
To illustrate, let us consider an one-loop triangular diagram
composed of
three electron propagators. We denote vertices of the diagram as $P$, $Q$
and $R$
respectively.
It is useful to recall that the
full
phase factor between two points $P$ and $Q$ is
\begin{eqnarray}
\Phi(P,Q) = \exp\left\{i\int^P_Q dx^\mu \left[A_\mu(x) +
{1\over{2}}F_{\mu\nu}(x- Q)^\nu\right]\right\}
\label{phasePQ}
\end{eqnarray} 
according to equation (\ref{phase}).
Here we use $P^{\mu}$ to denote the coordinate of the point $P$. Similarly
$Q^{\mu}$ and $R^{\mu}$ denote coordinates of the points $Q$ and $R$
respectively.
As discussed before, one can set $\Phi(P,Q)=1$ by choosing the
special
gauge
\begin{eqnarray}
{\bf A}_Q ({\bf x}) \equiv {\bf A}_0 ({\bf x})+\tilde {{\bf A}}_Q ({\bf x})
\end{eqnarray} 
with ${\bf A}_0 ({\bf x}) = {B/2}\cdot (-x_2,x_1,0)$ and
$\tilde {{\bf A}}_Q ({\bf x}) =
{B/2}\cdot (Q_2,-Q_1,0)$.
Similarly, one can respectively set $\Phi(R,P)$ and $\Phi(Q,R)$
to unity by choosing the gauges
\begin{eqnarray}
{\bf A}_P ({\bf x}) &\equiv& {\bf A}_0 ({\bf x})+\tilde {{\bf A}}_P ({\bf x})
\nonumber \\
{\bf A}_R ({\bf x}) &\equiv&  {\bf A}_0 ({\bf x})+\tilde {{\bf A}}_R ({\bf x}),
\end{eqnarray} 
with
$\tilde{{\bf A}}_P ({\bf x}) =
{B/2}\cdot (P_2,-P_1,0)$ and
$\tilde{ {\bf A}}_R ({\bf x}) =
{B/2}\cdot (R_2,-R_1,0)$ respectively. Apparently, $A_Q({\bf x})$,
$A_P({\bf x})$, and $A_R({\bf x})$ are distinct from one another.
Hence they cannot be
adopted simultaneously to set all phase factors to unity. In other words,
the phase factors
shall give rise to a non-trivial contribution to the three-point amplitude.
In fact this non-trivial contribution can be understood in an
alternative view.
Taking Eq. (\ref{phasePQ}) as an example, the integrand on the r.h.s.
of the equation
can be written as
${\cal A} \equiv A+{1\over {2}}F\Delta x$
where ${\cal A}\equiv
{\cal A}_{\mu}
dx^{\mu}$, $A\equiv A_{\mu} dx^{\mu}$, and
$F\Delta x\equiv F_{\mu\nu} dx^{\mu} (x-
Q)^\nu$
are all one-form. One can easily show that
${\cal A}$ is a closed form, i.e.,
\begin{eqnarray}
d{\cal A} =0.
\end{eqnarray} 
Note that ${\cal A}$ is exact if the first homology group 
is trivial, namely, $H_1(M)=0$. To be more specific,
if the gauge function $A_{\mu}(x)$ is regular everywhere,
then the one-form
${\cal A}$ is also regular. Therefore there exists a zero-form $\omega$ such that
${\cal A}=d\omega$ is an exact form. As a result, the
line integration which defines $\Phi(P,Q)$ is path independent.
In our problem, we need to compute the product of three phases:
$\Phi(P,Q)\cdot \Phi(R,P)\cdot \Phi(Q,R)$. 
It is then important to note 
that the one-form ${\cal A}$ in each of the above phases
depends on the boundary point of the path, despite the fact that
the gauge function $A_{\mu}(x)$ is regular.   
In other words, the gauge of ${\cal A}$ is chosen differently in each path, which
then gives rise to a non-trivial phase for a three-point amplitude. 
Precisely one may isolate the boundary dependencies of ${\cal A}$ 
by writing, for example,
${\cal A}={\cal A'}-F_{\mu\nu}dx^{\mu}Q^{\nu}$
in the case of $\Phi(P,Q)$. Apparently, ${\cal A'}$ is
an exact form universal to the three phases, while
$F_{\mu\nu}dx^{\mu}Q^{\nu}$ depends on the boundary point $Q$.
Using this separation, one may rewrite
each phase as
\begin{eqnarray}
\Phi (x,x') = \exp \left[ie \int^x_{x'}d\xi^\mu (A_\mu +
{1\over{2}}F_{\mu\nu} x^{\nu} )\right] \exp\left[-i{e\over{2}}
F_{\mu\nu}\int^x_{x'}d\xi^\mu x'^\nu \right].
\label{sep}
\end{eqnarray} 
Let us denote the first factor $\exp \left[ie \int^x_{x'}d\xi^\mu (A_\mu +
{1\over{2}}F_{\mu\nu} x^{\nu} )\right]$ as
$\Phi'(x,x')$.
Since $\Phi'(P,Q)\cdot \Phi'(R,P)\cdot \Phi'(Q,R)=\Phi'(Q,Q)=1$,
we conclude from Eq. (\ref{sep}) that
\begin{eqnarray}
\Phi(P,Q)\cdot \Phi(R,P)\cdot \Phi(Q,R)= \exp\left[-{ie\over 2}
(R-P)^{\mu}F_{\mu\nu}(P-Q)^{\nu}\right]. \label{phase1}
\end{eqnarray} 
This is the nontrivial phase contribution one must attach to
the amplitude of a three-point process when we write all weak field charged propagator
according to Eq. (\ref{weakB}).


\section{Applications}
\subsection{$\gamma\gamma\to \nu\bar{\nu}$}
The weak-field expansion derived in the last section has been applied to
calculate the amplitudes of $\gamma\gamma\to
\nu\bar{\nu}$  and its crossed processes in a homogeneous magnetic field 
less than 
$B_c$\cite{CTAS}.
According to the 
discussion in the previous section, 
the magnetic-field dependencies of above amplitudes 
reside in two places:
the first place is in the electron propagator which is
affected by the external magnetic field, while
the second place is in the overall phase which is a function of
the field strength tensor $F_{\mu\nu}$. 
Let us now take $\gamma\gamma\to \nu\bar{\nu}$ as an example
for illustration. Since the incoming photon energies are much less than
$m_W$, we can calculate the scattering amplitudes using the following 
effective four-fermion
interactions between leptons and neutrinos:
\begin{equation}
{\cal L}=-{G_F\over
\sqrt{2}}\left(\bar{\nu}_l\gamma_{\alpha}(1-\gamma_5)
\nu_l\right)\left(\bar{e}\gamma^{\alpha}(g_V-g_A\gamma_5)
e\right),
\end{equation}
where $g_V=1/2+2\sin^2 \theta_w$ and $g_A=1/2$ for $l=e$;
$g_V=-1/2+2\sin^2 \theta_w$ and $g_A=-1/2$
for $l=\mu,\tau$. The Feynman diagram contributing to  
$\gamma\gamma\to \nu\bar{\nu}$ is shown in Fig. 1 of 
Ref.\cite{CTAS}. 
We should remark that 
the contribution due to
$g_A$ is proportional to the neutrino mass in the limit of vanishing
magnetic field.
At $O(eB)$ in the limit $B\ll B_c$,
it gives no contribution to the
amplitude by the charge
conjugation invariance. Therefore we shall neglect
the contribution by $g_A$. Likewise, we shall also neglect contributions
by $g_V$ for $l=\mu,\tau$, since $-1/2+2\sin^2 \theta_w=0.04\ll 1$.

To $O(eB)$, the amplitude for $\gamma\gamma\to \nu\bar{\nu}$
can be written as $M\equiv M_1+M_2$,
where $M_1$ arises from inserting the external magnetic field to electron
propagators according to Eq. (\ref{weakB}), whereas
$M_2$ comes from expanding the overall phase factor for the
three-point function as shown in Eq. (\ref{phase1}).
Therefore one has
\begin{eqnarray}
M_1 &=&i4\pi\alpha eB {G_Fg_V\over \sqrt{2}}\bar u (p_2)
\gamma^\alpha (1-\gamma_5)v(p_1) \epsilon_1^\mu \epsilon_2^\nu \int
{d^4 l\over{(2 \pi)^4}} \nonumber \\
&\times& {\rm tr}\Bigg\{\gamma_\alpha (1-\gamma_5)\Bigg[
{\gamma_1\gamma_2[\gamma \cdot (l+k_1)_\parallel+m]\over{[(l+k_1)^2-m^2]^2}}
\gamma^\mu {i(\not{\! l}+m)\over{l^2-m^2}}\gamma^\nu
{i(\not{\! l}-\not{\! k_2}+m)\over{(l-k_2)^2-m^2}}\nonumber \\
&&~~~~~~~~~~~~~~~~+{i(\not{\! l}+\not{\! k_1}+m)
\over{(l+k_1)^2-m^2}}\gamma^\mu
{\gamma_1\gamma_2[\gamma \cdot l_\parallel+m]\over{(l^2-m^2)^2}}\gamma^\nu
{i(\not{\! l}-\not{\! k_2}+m)\over{(l-k_2)^2-m^2}}\nonumber \\
&&~~~~~~~~~~~~~~~~+{i(\not{\! l}+\not{\! k_1}+m)\over{(l+k_1)^2-m^2}}\gamma^\mu
{i(\not{\! l}+m)\over{l^2-m^2}}\gamma^\nu {\gamma_1\gamma_2[\gamma \cdot
(l-k_2)_\parallel+m]\over{[(l-k_2)^2-m^2]^2}} \Bigg]\Bigg\}, \label{M1}
\end{eqnarray} 
where $g_V=1-(1-4\sin^2 \theta_w)/2$ for $\nu_e$
and $m$ is the mass of the electron. The first and second term in
$g_V$ are the contributions from the $W$ and $Z$ exchanges, respectively.
To write down the amplitude $M_2$, we recall Eq. (\ref{phase1})
which states that the overall phase factor for $\gamma\gamma\to
\nu\bar{\nu}$ is
\begin{eqnarray}
\Phi(X,Y)\cdot \Phi(Z,X)\cdot \Phi(Y,Z)= \exp\left[-{ie\over 2}
(Z-X)^{\mu}F_{\mu\nu}(X-Y)^{\nu}\right]. \label{nunuph}
\end{eqnarray} 
With ${\bf B}$ in the forward $z$ direction and choosing
$X_{\mu}=(0,0,0,0)$, we arrive at
\begin{eqnarray}
\Phi(X,Y)\cdot \Phi(Z,X)\cdot \Phi(Y,Z)&=&
\exp\left\{({-iB\over{2}})(Y_1 Z_2-Y_2 Z_1)\right\} \nonumber \\
&\simeq& 1-{ieB\over{2}}(Y_1 Z_2-Y_2 Z_1) .\label{exphase}
\end{eqnarray} 
Since we calculate the amplitude only to $O(eB)$,
the first term
of the above expansion gives rise to $M_1$; while the second term
gives rise to $M_2$ which reads:
\begin{eqnarray}
M_2 &=&i4\pi\alpha eB {G_Fg_V\over \sqrt{2}}
\int d^4 Y~d^4 Z {-i\over{2}}(Y_1 Z_2-Y_2 Z_1) \int{d^4 l~d^4
q~d^4 r\over{(2 \pi)^{12}}}\epsilon^\mu_1 \epsilon^\nu_2 \nonumber \\
&\times& \exp[{-i(q-l-k_1)\cdot Y}] \exp[{-i(r-q-k_2)\cdot Z}]\bar u (p_2)
\gamma^\alpha (1-\gamma_5)v(p_1)\nonumber \\
&\times& {\rm tr}\left\{\gamma_\alpha (1-\gamma_5) {i(\not{\! l}+m)
\over{l^2-m^2}}
(-ie\gamma_\mu){i(\not{\! q}+m)\over{q^2-m^2}}(-ie\gamma_\nu)
{i(\not{\! r}+m)\over{r^2-m^2}}\right\}.
\end{eqnarray} 
We can recast the amplitude $M_2$ by using the equations
\begin{eqnarray}
Y_i\, \exp[{-i(q-l-k_1)\cdot Y}] &=& -
i{\partial\over{\partial l_i}}\exp[{-i(q-l-k_1)\cdot Y}],\nonumber \\
Z_i\, \exp[{-i(r-q-k_2)\cdot Z}] &=&
i{\partial\over{\partial r_i}}\exp[{-i(r-q-k_2)\cdot Z}],\nonumber
\end{eqnarray} 
and the integration by part, such that
\begin{eqnarray}
M_2 &=& i4\pi\alpha eB {G_Fg_V\over \sqrt{2}} \int d^4 Y~d^4 Z {-i\over{2}}
\int{d^4 l~d^4 q~d^4 r\over{(2 \pi)^{12}}}\epsilon^\mu_1
\epsilon^\nu_2\nonumber \\
&\times&\exp[{-i(q-l-k_1)\cdot Y}] \exp[{-i(r-q-k_2)\cdot Z}]
\bar u (p_2) \gamma^\alpha
(1-\gamma_5)v(p_1)\nonumber \\
&\times& \left[{\partial\over{\partial l_1}}
{\partial\over{\partial r_2}}-{\partial\over{\partial l_2}}
{\partial\over{\partial r_1}} \right]{\rm tr}\left\{\gamma_\alpha (1-\gamma_5)
{i(\not{\! l}+m)\over{l^2-m^2}}(-ie\gamma_\mu){i(\not{\! q}+m)\over{q^2-m^2}}
(-ie\gamma_\nu){i(\not{\! r}+m)\over{r^2-m^2}}\right\}.
\end{eqnarray}
Before we proceed to compute $M_1$ and $M_2$, 
we wish to reiterate the validity of
the above expansion. As we have pointed out in Ref. \cite{CTAS} that,
by dimensional analysis, any
given power of $eB$ in the expansion of $M$ is accompanied by
an equal power of $1/m^2$ (for $m >p$) or $1/p^2$ (for
$p> m$). Here $p$ denotes the typical
energy scale of external particles. Therefore, both $eB/m^2$ and $eB/p^2$ 
are much smaller than unity for $B\ll B_c\equiv m^2/e$.
Now performing the integration in $M_1$ and $M_2$, we obtain
\begin{eqnarray}
M\equiv M_1+M_2= {G_F g_V \alpha^{3/2}\over{\sqrt{2}\sqrt{4 \pi}}}\bar u (p_2)
\gamma_\alpha (1-\gamma_5)v(p_1) J^\alpha , \label{amplitude1}
\end{eqnarray} 
where\cite{private} 
\begin{eqnarray}
J^\alpha &=& C_1 (\epsilon_1 F \epsilon_2)(k^\alpha_1-k^\alpha_2)\nonumber \\
&+& C_2  [(\epsilon_1 F k_1) (k_1\cdot \epsilon_2) k^\alpha_2+(\epsilon_2 F
k_2)
(k_2\cdot \epsilon_1) k^\alpha_1]\nonumber \\
&+& C_3 [(\epsilon_1 F k_1) \epsilon_2^\alpha+(\epsilon_2 F k_2)
\epsilon_1^\alpha]\nonumber \\
&+& C_4  [(\epsilon_1 F k_2) (k_1\cdot \epsilon_2) k^\alpha_1+(\epsilon_2 F
k_1)
(k_2\cdot \epsilon_1) k^\alpha_2]\nonumber \\
&+& C_5  [(\epsilon_1 F k_2) (k_1\cdot \epsilon_2) k^\alpha_2+(\epsilon_2 F
k_1)
(k_2\cdot \epsilon_1) k^\alpha_1]\nonumber \\
&+& C_6 [(\epsilon_1 F k_2) \epsilon_2^\alpha+(\epsilon_2 F k_1)
\epsilon_1^\alpha]\nonumber \\
&+& C_7 (k_2\cdot \epsilon_1)(k_1\cdot \epsilon_2)[(Fk_1)^\alpha+(Fk_2)^\alpha]
\nonumber \\
&+& C_8 (\epsilon_1\cdot \epsilon_2) [(Fk_1)^\alpha+(Fk_2)^\alpha]\nonumber \\
&+& C_9 (k_1 F k_2) (\epsilon_1\cdot \epsilon_2) (k_1^\alpha-k_2^\alpha)
\nonumber \\
&+& C_{10} (k_1 F k_2) (k_2\cdot \epsilon_1)(k_1\cdot \epsilon_2)
(k_1^\alpha-k_2^\alpha) \nonumber \\
&+& C_{11} (k_1 F k_2) [(k_2\cdot \epsilon_1)
\epsilon_2^\alpha-(k_1\cdot \epsilon_2) \epsilon_1^\alpha]
\end{eqnarray} 
with, for instance, $(\epsilon_1F\epsilon_2)\equiv \epsilon_1^{\mu}F_{\mu\nu}
\epsilon_2^{\nu}$ and $(Fk_1)^{\alpha}\equiv F^{\alpha\beta}k_{1\beta}$.
The coefficient functions $C_i$'s are given as follows:
\begin{eqnarray}
C_1&=& -{8\over{m^2}} \Big(I[0, 0, 1] + I[0, 0, 2] -
4 I[1, 1, 1] - 5 I[1, 1, 2]
+ 2 I[2, 1, 1] + 2 I[2, 1, 2]\nonumber \\
&&~~~+ t I[2, 1, 2] + 2 I[2, 2, 1] + 2 I[2, 2, 2] - 5 t I[3, 2, 2]
+ 2 t I[4, 2, 2] + 2 t I[4, 3, 2]\Big), \nonumber\\
C_2&=& -{8\over{m^4}} \Big(I[1, 1, 2] - 2 I[2, 1, 2] - 3 I[2, 2, 2]
+ 4 I[3, 2, 2] + 2 I[3, 3, 2] - 4 I[4, 3, 2]\Big), \nonumber \\
C_3&=& -{4\over{m^2}} \Big(2 I[0, 0, 2] - 4 I[1, 1, 1] - 4 I[1, 1, 2]
- t I[1, 1, 2] + 2 c I[2, 1, 2] + 2 I[2, 2, 1]\nonumber \\
&&~~~ + 2 I[2, 2, 2] + 3 t I[2, 2, 2] - 4 t I[3, 2, 2] - 2 t I[3, 3, 2]
+ 2 t I[4, 3, 2]\Big),\nonumber \\
C_4&=& -{16\over{m^4}} \Big(5 I[3, 2, 2] - 2 I[4, 2, 2] - 4 I[4, 3, 2] \Big),
\nonumber \\
C_5&=& -{8\over{m^4}} \Big(I[1, 1, 2] + 2 I[2, 1, 2] - I[2, 2, 2]
- 10 I[3, 2, 2] + 8 I[4, 2, 2] + 4 I[4, 3, 2]\Big),\nonumber \\
C_6&=& -{4\over{m^2}} \Big(2 I[0, 0, 1] + 2 I[0, 0, 2] - 4 I[1, 1, 1]
- 4 I[1, 1, 2] - t I[1, 1, 2] \nonumber \\
&&~~~- 4 I[2, 1, 1] - 4 I[2, 1, 2]- 2 I[2, 2, 1] - 2 I[2, 2, 2]
+ t I[2, 2, 2] \nonumber \\
&&~~~+ 2 t I[3, 2, 2]- 4 t I[4, 2, 2] -2 t I[4, 3, 2]\Big),\nonumber \\
C_7&=& {8\over{m^4}} \Big(I[1, 1, 2] - 2 I[2, 1, 2] - I[2, 2, 2]
+ 4 I[3, 2, 2] -4 I[4, 3, 2]\Big), \nonumber \\
C_8&=& {4\over{m^2}} \Big(2 I[0, 0, 2] - 4 I[1, 1, 1] - 4 I[1, 1, 2]
- t I[1, 1, 2] + 2 t I[2, 1, 2] + 2 I[2, 2, 1]\nonumber \\
&&~~~ + 2 I[2, 2, 2] + t I[2, 2, 2] - 4 t I[3, 2, 2] +2 t I[4, 3, 2]\Big),\nonumber\\
C_9&=& -{8\over{m^4}} \Big(I[1, 1, 2] + 2 I[2, 1, 2] + 4 I[2, 1, 3]
- I[2, 2, 2] - 10 I[3, 2, 2] - 12 I[3, 2, 3] \nonumber \\
&&~~~+ 4 I[4, 2, 2] +4 I[4, 2, 3] + 4 t I[4, 2, 3] + 4 I[4, 3, 2]
+4 I[4, 3, 3] - 12 t I[5, 3, 3] \nonumber \\
&&~~~+4 t I[6, 3, 3] + 4 t I[6, 4, 3]\Big),\nonumber \\
C_{10}&=& {64\over{m^6}} \Big(I[4, 2, 3] - 4 I[5, 3, 3] + 2 I[6, 3, 3]
+ 2 I[6, 4, 3]\Big),\nonumber \\
C_{11}&=& -{8\over{m^4}} \Big(I[1, 1, 2] + 2 I[2, 1, 2] + 4 I[2, 1, 3]
- I[2, 2, 2] - 4 I[3, 2, 3] - 4 I[4, 2, 2]\nonumber \\
&&~~~ - 4 I[4, 2, 3] - 4 I[4, 3, 2] - 4 I[4, 3, 3] + 4 t I[5, 3, 3]
- 4 t I[6, 3, 3] - 4 t I[6, 4, 3]\Big),
\end{eqnarray} 
where
\begin{eqnarray}
I[a,b,c] \equiv \int^1_0 dx\int^{1-x}_0 dy
{x^b y^{a-b}\over{(1-txy-i\varepsilon)^c}}
\end{eqnarray} 
with $t\equiv 2 {k_1\cdot k_2\over{m^2}}$.
Our result is an extension of the calculation in Ref. \cite{Sha}
which considers only the low energy limit $k_1\cdot k_2 \ll m^2$.
In such a limit, one can calculate $M$ using the effective Lagrangian
for $\gamma\gamma\to \nu\bar{\nu}\gamma$\cite{DR} and replacing one of
the photons by the external magnetic field.

With the amplitude $M$, it is straightforward to compute 
the scattering cross section $\sigma_B (\gamma\gamma\to\nu\bar\nu)$.         
in the background magnetic field
Since $\gamma\gamma\to \nu\bar{\nu}$ could contribute to the
energy-loss of a magnetized star, it is useful to compute the
stellar energy-loss rate $Q$, which is related to $\sigma_B$
through\cite{COMM}
\begin{eqnarray}
Q={1\over{2 (2\pi)^6}}\int {2d^3 \vec k_1\over{e^{\omega_1/T}-1}}\int
{2d^3 \vec k_2\over{e^{\omega_2/T}-1}}{(k_1\cdot k_2)\over{\omega_1\omega_2}}
(\omega_1+\omega_2)\sigma_B (\gamma\gamma\to\nu\bar\nu).        \label{QQ}
\end{eqnarray} 
In Ref. \cite{Sha}, $Q$ is calculated based upon an
approximated cross section obtained in the limit $E_{\gamma}\ll m$.
Such a
calculation is repeated in our earlier work \cite{CTAS} which is
based upon the cross section $\sigma_B(\gamma\gamma\to
\nu\bar{\nu})$ obtained from the amplitude $M$ in 
Eq. (\ref{amplitude1}).
We found that, for temperatures below
$0.01$ MeV, the effective-Lagrangian
approach employed in Ref.\cite{Sha}works very well. On the other
hand, this approach becomes rather inaccurate for temperatures greater
than $1$ MeV. At $T=0.1$ MeV, our calculation gives an
energy-loss
rate almost two orders of magnitude greater than the result from the
effective Lagrangian. Such a behavior can be understood from the
energy dependence of the scattering cross section, as shown in Fig. 2
of Ref. \cite{CTAS}.
It is clear that, for $T=0.1$ MeV, $Q$ must have received
significant contributions from scatterings with
$\omega\approx m$. At this energy, the full calculation gives a much
larger
scattering cross section than that given by the effective Lagrangian.
By comparing the predictions of the full calculation and
the effective-Lagrangian approach \cite{Sha}, we conclude that
the applicability of the latter to the energy-loss rate
is quite restricted. While the effective Lagrangian works reasonably
well
with $\omega< \,  0.1 m $ , it would give a poor approximation on $Q$
unless $T < \, 0.01 m$.

\subsection{$\gamma\to \nu\bar{\nu}$, $\nu\to \nu\gamma$}
In order to compare our approach with previous ones, we consider the
simple two-body decay modes 
$\gamma\to \nu\bar{\nu}$\cite{GN,SK,DMH} and $\nu\to \nu\gamma$\cite{IR}
in a background magnetic field.
We shall limit the energies of incoming particles to be less than the
pair-production threshold $2m$. For if $E_{\gamma}, E_{\nu}> 2m$,
the dominant decay modes should become $\gamma \to e^+ e^-$ and
$\nu\to \nu e^+ e^-$ respectively. For incoming energies below the 
pair-production threshold, 
it turns out that the photon momenta in both  
$\gamma\to \nu\bar{\nu}$ and $\nu\to \nu\gamma$
are space-like\cite{adler}. Hence the former process is kinematically 
forbidden. The amplitude of the latter process can be written as
\begin{equation}
{\cal M}(\nu(p_1)\to \nu(p_2)\gamma(q))
=-{G_F\over \sqrt{2}e}Z\epsilon^{\alpha}\bar{u}(p_2)
\gamma^{\beta}(1-\gamma_5)
u(p_1) 
\left(g_V\Pi_{\alpha\beta}(-q)-g_A\Pi^5_{\alpha\beta}(-q)\right),
\label{chek}
\end{equation}  
where $\Pi_{\alpha\beta}$ and $\Pi^5_{\alpha\beta}$ are vector-vector
and vector-axial vector two-point functions given by
\begin{eqnarray}
\Pi_{\alpha\beta}(q)&=&-e^2\int {d^4k\over (2\pi)^4}Tr\left[\gamma_{\alpha}
{\cal G}(k-q)\gamma_{\beta}{\cal G}(k)\right]\nonumber , \\
\Pi^5_{\alpha\beta}(q)&=&-e^2\int {d^4k\over (2\pi)^4}Tr\left[\gamma_{\alpha}
{\cal G}(k-q)\gamma_{\beta}\gamma_5{\cal G}(k)\right].
\end{eqnarray}
The factor $Z$ is the wave-function renormalization
constant of the photon field, induced by the effect of external magnetic 
fields. Since the deviation of $Z$ from the unity is rather small, 
proportional to
the fine structure constant $\alpha$, we shall set $Z=1$ in our subsequent 
discussions.  

The structures of the two-point functions 
$\Pi_{\alpha\beta}$ and $\Pi^5_{\alpha\beta}$ 
were given in previous literature\cite{TSAI,DMH}
\begin{eqnarray}
\Pi_{\alpha\beta}(q)&=&A\left(q_{\parallel}^2g_{\parallel\alpha\beta}-
q_{\parallel\alpha}q_{\parallel\beta}\right)
+B\left(-q_{\bot}^2g_{\bot\alpha\beta}-
q_{\bot\alpha}q_{\bot\beta}\right)
+C\left(q^2g_{\alpha\beta}-
q_{\alpha}q_{\beta}\right)\nonumber ,\\
\Pi^5_{\alpha\beta}(q)&=&C_{\parallel}\left(q_{\parallel}^2\tilde{F}_
{\alpha\beta}
+q_{\parallel\alpha}(\tilde{F}q)_{\beta}+q_{\parallel\beta}
(\tilde{F}q)_{\alpha}\right)\nonumber \\
&+&C_{\bot}\left(-q_{\bot}^2\tilde{F}_{\alpha\beta}
+q_{\bot\alpha}(\tilde{F}q)_{\beta}+q_{\bot\beta}
(\tilde{F}q)_{\alpha}\right).
\label{structure}
\end{eqnarray}      
We wish to remind the reader that $q_{\bot}^2=(q^1)^2+(q^2)^2$ for a 
magnetic field in the $+z$ direction. The calculations of 
$\Pi_{\alpha\beta}$ and
$\Pi^5_{\alpha\beta}$ for $B<B_c$ are straightforward using the 
weak field expansion derived in 
Eqs. (\ref{weakB}), (\ref{B23}), and (\ref{B45}). Due to
charge-conjugation and gauge invariances, the magnetic-field effects to 
$\Pi_{\alpha\beta}$ and $\Pi^5_{\alpha\beta}$ begin at the order 
$e^2B^2$ and $e^3B^3$ respectively. The subleading 
contributions are then of the 
order $e^4B^4$ and $e^5B^5$ respectively. 
The coefficient functions of $\Pi_{\alpha\beta}(q)$ are given 
by
\begin{eqnarray}
A&=&{i\alpha\over \pi}\left[-{7\over 45}
({B\over B_c})^2+({26\over 315}-{52\over 945}{\omega^2
\over m^2}\sin^2\theta)({B\over B_c})^4+\cdots\right],\nonumber \\
B&=&{i\alpha\over \pi}\left[{4\over 45}
({B\over B_c})^2+(-{16\over 105}+{4\over 135}{\omega^2
\over m^2}\sin^2\theta)({B\over B_c})^4+\cdots\right],\nonumber \\
C&=&{i\alpha\over \pi}\left[({2\over 45}-{1\over 105}{\omega^2
\over m^2}\sin^2\theta)({B\over B_c})^2\right.\nonumber \\
&+&\left.(-{4\over 105}+{44\over 1575}{\omega^2
\over m^2}\sin^2\theta-{10\over 2079}{\omega^4
\over m^4}\sin^4\theta)({B\over B_c})^4+\cdots\right],
\label{VV}
\end{eqnarray}
where $\omega$ is the photon energy while $\theta$ is the angle between the
the magnetic-field direction and the direction of photon propagation. 
For the coefficient functions of
$\Pi^5_{\alpha\beta}(q)$, we find
\begin{eqnarray}
C_{\parallel}&=&{i\alpha\over B\pi}\left[{1\over 70}
{\omega^2
\over m^2}\sin^2\theta({B\over B_c})^3
+(-{26\over 945}{\omega^2
\over m^2}\sin^2\theta+{10\over 693}{\omega^4
\over m^4}\sin^4\theta)({B\over B_c})^5+\cdots\right],\nonumber \\
C_{\bot}&=&{i\alpha\over B\pi}\left[(-{1\over 15}+{1\over 70}{\omega^2
\over m^2}\sin^2\theta)({B\over B_c})^3\right.\nonumber \\
&+&\left.({8\over 63}-{86\over 945}{\omega^2
\over m^2}\sin^2\theta+{10\over 693}{\omega^4
\over m^4}\sin^4\theta)({B\over B_c})^5+\cdots\right].
\label{AV}
\end{eqnarray}
It should be noted that the validity 
of weak-field expansion in Eqs.(\ref{VV}) and (\ref{AV})
also depends on the ratio 
$r\equiv \omega^2\sin^2\theta B^2/ m^2B_c^2$, 
besides the requirement $({B\over B_c})^2\ll 1$. For a  sufficiently large   
photon energy such that $r>1$, the expansion in Eqs. (\ref{VV}) and (\ref{AV}) 
may break down. However, since we have limited the photon energy
to $\omega< 2m$, the ratio $r$ is automatically smaller than 1.

The computation of   
$\nu\to \nu\gamma$ width requires the knowledge of
photon index of refraction 
$n\equiv {\vert \vec{q}\vert\over \omega}$.
The index of refraction can be calculated 
from the two-point
function $\Pi_{\alpha\beta}(q)$. It is well known  that 
$n$ depends on the photon polarizations.
For the magnetic field in the 
$+z$ direction, the polarization states with distinct index of refraction
are $\epsilon_{\bot}^{\mu}=(0,0,1,0)$ and $\epsilon_{\parallel}^{\mu}
=(0,-\cos\theta,0,\sin\theta)$. Here we have adopted the convention
that $q^{\mu}=(\omega,\omega\sin\theta,0,\omega\cos\theta)$, i.e., photon
propagates on the $x-z$ plane with an angle $\theta$ to the 
magnetic field direction. Hence $\vec{\epsilon}_{\bot}$ is the polarization 
vector perpendicular to the $x-z$ plane 
while $\vec{\epsilon}_{\parallel}$ lies on the
$x-z$ plane. The photon dispersion relation is given by
\begin{equation}
q^2-i\Pi_{a}=0,
\label{DIS}
\end{equation}
where $\Pi_{a}=\epsilon^{\alpha}_{a}
\Pi_{\alpha\beta}\epsilon^{\beta}_{a}$. Here the index $a$ stands for the
polarization states, namely $a= \bot$ or $\parallel$.
Combining Eqs. (\ref{VV}) and (\ref{DIS}), we arrive at 
\begin{eqnarray}
\left[1+iB\sin^2\theta+iC\right]q^2&=&iB\omega^2\sin^2\theta, 
\,\nonumber \\
\left[1+iC\right]
q^2&=&-iA\omega^2\sin^2\theta,\, 
\end{eqnarray}
for polarization states $a= \bot$ and $a=\parallel$ respectively.
Since the electomagnetic coupling constant is rather small, the left hand 
side of the above equations may be approximated by $q^2$. Using the 
definition $q^2=\omega^2\cdot(1-n^2)$, we obtain
\begin{eqnarray}
n_{\bot}&=&1+{\alpha\over \pi}\left[{2\over 45}({B\over B_c})^2
+(-{8\over 105}+{2\over 135}{\omega^2\over m^2}\sin^2\theta)
({B\over B_c})^4+\cdots\right]\sin^2\theta,\nonumber\\
n_{\parallel}&=&1+{\alpha\over \pi}\left[{7\over 90}({B\over B_c})^2
+(-{13\over 315}+{26\over 945}{\omega^2\over m^2}\sin^2\theta)
({B\over B_c})^4+\cdots\right]\sin^2\theta.
\label{refr}
\end{eqnarray}
It is seen that the leading contributions to $n_{\bot}$ and 
$n_{\parallel}$ agree with the results obtained by Adler\cite{adler}.
The next-to-leading contribution to $n_{\bot,\parallel}$ depend on 
both the photon energy $\omega$ and 
the photon propagation direction\cite{comm}.

Given the above photon dispersion relation, we proceed to compute
the $\nu\to \nu\gamma$ width in the sub-critical background magnetic field.
We note that the most recent calculation of 
$\nu\to \nu\gamma$ width is performed by Ioannisian and Raffelt\cite{IR}. 
Following their approach, we write the width of this process as
\begin{equation}
\Gamma={1\over 16\pi E_1^2}\int_0^{\omega_{\rm max}}d\omega\sum_{\rm pols}
{\vert {\cal M}\vert}^2,
\label{width}
\end{equation}
where ${\cal M}$ is the amplitude given by Eq. (\ref{chek}), 
$E_1$ is the neutrino energy, and 
$\omega_{\rm max}={\rm min}
(E_1,\omega_c)$ 
with $\omega_c$ the critical photon energy beyond which the photon four 
momentum becomes time-like and the Cherenkov condition no longer holds. 
We note that, in deriving the above width, one has taken 
the collinear approximation that the particles in the
initial and final states are all parallel with one another. 
The correction to such an approximation is small, proportional to
the fine structure constant $\alpha$. From Eq. (\ref{chek}), we obtain
\begin{equation}
{\vert {\cal M}\vert}^2={4g_A^2G_F^2\over e^2}\epsilon^{\alpha}(a)
\epsilon^{*\alpha'}(a)\left(p_1^{\beta}p_2^{\beta'}+
p_1^{\beta'}p_2^{\beta}\right)\Pi^5_{\alpha\beta}\Pi^5_{\alpha'\beta'},
\label{square}
\end{equation}
where $a=\parallel, \bot$. The contribution by the vector-vector
two-point function $\Pi_{\alpha\beta}$ is negligible since both
$p_1$ and $p_2$ are approximately parallel to the photon momentum
$q$ and $q^{\beta}\Pi_{\alpha\beta}=0$ due to the gauge invariance.         
The fact that both $p_1$ and $p_2$ are approximately parallel to $q$
also has a consequence on the polarization dependencies of 
${\vert {\cal M}\vert}^2$. This is easily seen with
\begin{equation}
q^{\beta}\Pi^5_{\alpha\beta}=2\left(C_{\parallel}q_{\parallel}^2
-C_{\bot}q_{\bot}^2\right)(\tilde{F}q)_{\alpha} .
\end{equation}
For a $B$ field in the $+z$ direction, $(\tilde{F}q)_{\alpha}$ is 
nonvanishing only for $\alpha=0,3$. Given 
$\epsilon_{\bot}^{\mu}=(0,0,1,0)$ and $\epsilon_{\parallel}^{\mu}
=(0,-\cos\theta,0,\sin\theta)$ as stated earlier, one immediately see that
${\vert {\cal M}\vert}^2$ vanishes for a photon in a 
$\bot$ mode. Hence the photon radiated from the neutrino is 
polarized, with its 
polarization vector lying on the surface 
spanned by $\vec{q}$ and $\vec{B}$. 

The width of $\nu\to \nu\gamma$ can be readily calculated using 
Eqs. (\ref{width}),
(\ref{square}) and (\ref{structure}). We have
\begin{equation}
\Gamma={g_A^2G_F^2B^2\over 2\pi^2E_1\alpha}
\sin^6\theta\int_0^{\omega_{\rm max}}
d\omega (E_1-\omega)\omega^4{\vert C_{\parallel}-C_{\bot} \vert}^2.
\end{equation}
Since $C_{\parallel}$ and $C_{\bot}$ are already given by Eq. (\ref{AV}), 
$\Gamma$ can be 
easily determined once $\omega_{\rm max}$ is specified. 
Since $E_1< 2m$, which implies $\omega< 2m$,
the photon refractive index is always greater than 1
as indicated by Eq. (\ref{refr}). Hence the critical 
energy $\omega_c$ for photon dispersion relation to cross the light cone 
is greater than $2m$. Thus $\omega_{max}\equiv{\rm min}(E_1,\omega_c)=
E_1$. The width $\Gamma$ is given as follows:
\begin{equation}
\Gamma={2G_F^2\alpha E_1^5\over 135(2\pi)^4}\sin^6\theta\left
[{1\over 50}({B\over B_c})^6-\left({8\over 105}-{1\over 49}{E_1^2\sin^2\theta
\over m^2}\right)({B\over B_c})^8+\cdots \right] .
\end{equation}
Comparisons of our result with the earlier results of 
Ref.\cite{DMH,IR} are in order. First, we
focus on the weak field region $B< B_c$ while Refs. \cite{DMH,IR} considers the
general magnetic field and the corresponding coefficient functions 
$C_{\parallel,\bot}$ 
are expressed in double integrals. Second, 
due to a different convention, the coefficient functions obtained in
Ref.\cite{IR}, denoted as $C'_{\parallel,\bot}$, are related to ours via the
relation $\vert C_{\parallel}-C_{\bot}\vert={e^4\over 32\pi^2m^2}
\vert C'_{\parallel}-2C'_{\bot}\vert$ where
\begin{eqnarray}
C'_{\parallel}&=&im^2\int_0^{\infty}ds\int_{-1}^1 dve^{-is\phi_0}(1-v^2),
\nonumber \\
C'_{\bot}&=&im^2\int_0^{\infty}ds\int_{-1}^1 dve^{-is\phi_0}R,
\end{eqnarray}
with
\begin{eqnarray}
\phi_0&=&m^2+{1-v^2\over 4}q_{\parallel}^2+{\cos(eBsv)-\cos(eBs)\over
2eBs\sin(eBs)} q_\bot^2 ,\nonumber \\
R&=&{1-v\sin(eBsv)\sin(eBs)-\cos(eBs)\cos(eBsv)\over
\sin^2(eBs)}.
\end{eqnarray}    
To compare the two sets of results, it is useful to realize that
one can rotate the integration contour, $s\to -is$, in the above 
integrals, provided $q_0\equiv \omega
< 2m$. In this way, the phase $e^{-is\Phi_0}$ turns into
$e^{-s\Phi_0}$ and becomes highly suppressed 
for a large $s$. For $B<B_c$, such a behaviour permits one to simultaneously
perform the weak-field and low-energy expansions with respect to
$C_{\parallel,\bot}$. The results of expanding 
$\vert C'_{\parallel}-2C'_{\bot}\vert$
may be organized into the sum of the following series $\sum_{n=0}a_n
(\omega^2 \sin^2\theta/m^2)^n (B^2/B_c^2)^n$, $\sum_{n=0}b_n
(\omega^2 \sin^2\theta/m^2)^n (B^2/B_c^2)^{n+1}$, $\sum_{n=0}c_n
(\omega^2 \sin^2\theta/m^2)^n (B^2/B_c^2)^{n+2}
\cdots$. One observes that the
coefficients $a_1$ and $a_2$ correspond to the $O(\omega^2 B^2)$ and $O(\omega^4
B^4)$ terms in our $\vert C_{\parallel}-C_{\bot}\vert $ respectively. 
We found that all the coefficients $a_i's$ vanish. This is indeed reflected 
in our calculations where the $O(\omega^2 B^2)$ and $O(\omega^4
B^4)$ terms in $\vert C_{\parallel}-C_{\bot}\vert$ vanish as well. 
We also found agreements between the coefficients $b_{0,1}$ 
and the corresponding 
$O(B^2)$, $O(\omega^2 B^4)$ terms in $\vert C_{\parallel}-C_{\bot}\vert$.
Although we did not compare the coefficient $c_0$ with the $O(B^4)$ term
in $C_{\parallel}$, 
due to the growing complexity in computing the general    
coefficients $c_i's$, the agreements we just found with 
respect to the first two series
seems rather compelling. Due to these agreements, we also confirm the 
statement made in Ref.\cite{IR} that the earlier calculation on 
$\Pi^5_{\mu\nu}$ is incorrect.  

From the above comparisons, we have seen that
our approach, in spite of less general, is convenient for
obtaining the analytic amplitudes of physical  
processes in a sub-critical background magnetic field. 
In such a magnetic field, it suffices to know the leading
and sub-leading terms in the weak-field expansion. Our approach 
produces those terms directly from Feynman diagrams. 

The work on extending the present analysis to the more complicated processes,
such as the photon splitting $\gamma\to \gamma\gamma$ and
the pair production $\gamma\to e^+e^-$ is currently being 
pursued. For the latter process, we have exploited the 
analytical properties of the vacuum polarization function $\Pi^{\mu\nu}$
in the background magnetic field. For a 
sub-critical magnetic field, it is possible to obtain a simple 
expression for the absorption coefficient(the pair-production width)
for arbitrary photon energies\cite{NEW}. This is an 
improvement to the previous work where a simplified 
expression is possible only for $\omega \gg m$\cite{TSE}.
For the former process, $\gamma\to
\gamma\gamma$, our result shall serve as an additional check   
to the previous results\cite{adler,baier}.


\section{Conclusion}
In this paper, we have developed the weak-field expansion
technique for processes
occurring in a background magnetic field. This expansion is performed with
respect to internal electron propagators which are affected by the
background magnetic field.
In some processes, our approach is valid for general
external
momenta even if they are much greater than the electron mass $m$. 
For external momenta much greater than $m$,
the effective-Lagrangian approach is no longer appropriate.
To illustrate this point, we
calculated the amplitude  of
$\gamma\gamma \to \nu\bar{\nu}$ under a background magnetic field, and
consequently
determined the stellar energy-loss rate $Q$ due to this process.
It is interesting to find
that the effective-Lagrangian approach is inappropriate
for computing the stellar energy-loss rate due to
$\gamma\gamma \to \nu\bar{\nu}$, unless the star temperature is
less than $0.01\, m$. This result reflects clearly the importance of
our approach. In fact, our approach can be applied to many other processes.
In this regard, we also discussed the 
processes $\gamma\to \nu\bar{\nu}$ and 
$\nu\to \nu\gamma$ under a strong background magnetic field.
We found that the validity of weak-field expansion with respect to
the above processes are also determined by the parameter 
$r\equiv \omega^2\sin^2\theta B^2/m^2B_c^2$, besides the requirement
$B<B_c$. For energy below pair production threshold,
the parameter $r$ is less than 1, which causes no trouble to the weak-field
expansion.
We found that $\gamma\to \nu\bar{\nu}$ is kinematically forbidden 
while $\nu\to \nu\gamma$ is permitted by the phase space.
Our predictions on the latter process 
agrees with previous works\cite{IR}. It has also been
pointed out that
our approach, although less general, is convenient for
obtaining the analytic amplitudes of physical  
processes in a sub-critical background magnetic field.   

We are currently extending the weak-field expansion
technique to the photon splitting process 
$\gamma\to \gamma\gamma$\cite{adler,baier}
and the pair production
process $\gamma\to e^+e^-$. 
\cite{adler,TSE}. Both processes are
of great interests in the physics of pulsars on which the background magnetic fields
are close to the critical value $B_c$.

\acknowledgments

This work was supported in part by the National Science Council 
under the Grant Nos. NSC-89-2112-M-009-001, NSC-89-2112-M-009-035, and
NSC-89-2112-M001-001.



\begin{thebibliography}{99}
\bibitem{COMM} We have neglected the Pauli blocking factor for the final state
fermions. Adding these factors does not change our result significantly.
\bibitem{comm}
In principle, such a result can also be obtained 
from Eq. (51) of Ref.\cite{adler} by 
simultaneously performing the weak-field and low-energy expansions.
\end{thebibliography}
\end{document}